\documentclass[a4paper, preprint, preprintnumbers, superscriptaddress, showpacs, showkeys, floatfix, twocolumn, aps, pra, 10pt]{revtex4-2}

\usepackage[utf8]{inputenc}
\usepackage[english]{babel}
\usepackage{amssymb}
\usepackage{physics}
\usepackage{amsfonts}
\usepackage{amsmath}
\usepackage{mathtools}
\usepackage{bbm}
\usepackage{amsthm}
\usepackage{xcolor}
\usepackage{appendix}

\usepackage{tikz,tkz-euclide}
\usetikzlibrary{decorations.pathmorphing, decorations.pathreplacing, decorations.text, shapes, calc, arrows, positioning, through, patterns, backgrounds, 3d, scopes, decorations.pathreplacing, angles, quotes, fit, shadings, patterns.meta,
 quotes,arrows.meta}

\usepackage{orcidlink}

\begin{document}

\title{On the issues arising when defining an X gate for qudits: Extending the Bit-Flip channel to $d$-dimensional systems}

\author{Gómez, Jean F.\orcidlink{0009-0000-5781-4089}
}
\email{jfgomez@usb.ve}

\author{Albrecht, Hermann L.\orcidlink{0000-0002-5735-8340}
}
\email{albrecht@usb.ve}
\affiliation{Departamento de F\'{\i}sica, Universidad Sim\'on Bol\'{\i}var, AP 89000, Caracas 1080, Venezuela.}

\begin{abstract}
Given the current interest in quantum information tasks involving higher-dimensional systems, we discuss some of the issues that appear when extending the bit-flip channel to qutrit systems. The difficulties arise from different interpretations of the Pauli X gate for qubits, leading to at least three nonequivalent formulations. We compare our results with the commonly used cyclic one-parameter trit flip channels and demonstrate that they are particular cases of those more general formulations we present here. We also extend these channels to higher-dimensional qudit systems, defining various dit flip channels. Finally, we study their impact on the Negativity, as an entanglement measure, of qubit-qutrit and 2-qutrit Werner states. In doing so, we show the distinctiveness of these versions, as they affect the states’ entanglement in very distinct ways.
\end{abstract}
\preprint{SB/F/496-24}
\maketitle

Higher-dimensional quantum systems, usually labeled as qudits, have been found to help reduce the resources needed for quantum computing operations \cite{extending_frontier_qutrits, efficient_toffoli} compared with the more studied two-dimensional qubits. These $d$-level quantum systems have been shown to help optimize various quantum algorithms \cite{qudits_faster_algo}, and their study within quantum information science and technology has drawn attention within the scientific community in recent years. The qutrit, the simplest qudit after the qubit, has been widely used within this context \cite{Sri2026Qutrit-basedQC}.

A qubit gate commonly used in quantum computing and communications is the Pauli X gate \cite{Nielsen-QIT, Nakahara-QC}, which, in the context of noisy quantum mechanics, is implemented in the bit-flip channel \cite{electronics13020439}. In higher dimensional systems, the corresponding X gate is commonly defined in the literature as a forward cyclic permutation \cite{Patera1988PauliGates-Qudits-Shift, PhysRevA.70.012302, PhysRevA.99.023825}. Nevertheless, the richer structure of qudits allows a more open-ended interpretation of what a ``flip” means.

Unlike the depolarizing or amplitude-damping quantum channels \cite{Nielsen-QIT, Nakahara-QC}, extending the bit-flip to higher-dimensional systems is not straightforward due to the non-binary nature of qudits. In particular, for qutrit systems, multiple interpretations can be proposed, as it is not obvious how a trit can be flipped or negated. In the literature, some trit-flip channels have been proposed and studied \cite{ramzan2011effect, wei2013geometric, doustimotlagh2015quantum, jiang2024joint}, and the extension of Pauli gates to higher dimensional systems is still an active field of research \cite{Sri2026Qutrit-basedQC}.

As we previously stated, there are different meanings to the flipping of a qudit that are indistinguishable from each other in a two-dimensional system. For instance, we can interpret the quantum bit-flip operation as one where an element of the qubit basis is flipped similarly to the negation operation in classical computing. 

However, we can interpret its action on a qubit state as obtaining the corresponding orthogonal element to the given input basis state. Finally, we can also interpret it as a successor and a predecessor operation to its cyclic-ordered basis. All of the abovementioned operations are equivalent in two-dimensional systems but will have a different meaning in higher-dimensional spaces. 

In quantum error correction \cite{Roffe2019QuantumErrorCorrect, Deepika2026ReviewQuantErrCorr}, the bit-flip channel, phase-flip channel, and combinations of them are the types of errors mostly considered in this context. With the widening interest in qudit quantum computation, extending these quantum channels to higher dimensions is not only theoretically appealing but also a practical concern. Moreover, since such extensions broaden the possible noise channels involved in qudit flipping, their study also contributes to the development of practical applications, such as quantum error correction codes.

In this paper, we discuss three quantum channels that represent the different meanings of flipping qutrits. In doing so, we also discuss the issues and limitations of defining Pauli-type gates and channels in higher dimensions, as, for instance, the necessity of choosing which of the characteristics of qubit Pauli gates, such as them being unitary, traceless, and (skew-)symmetric, to maintain when extending them to qudits. Finally, we present a general pathway to extend them to higher dimensions.

We have organized this article as follows. In Sections \ref{sec:IDF} and \ref{sec:su(d)-IDF}, we discuss the extension of the bit-flip channel to higher-dimensional systems by considering the flipping of two basis elements while the remaining ones are invariant. In Section \ref{sec:IDF}, this formulation relies on unitary operators, while in Section \ref{sec:su(d)-IDF}, we use the corresponding generalized Gell-Mann matrices associated with the Pauli X gate. Then, in Section \ref{sec:SDF}, we define the qudit-flip channel as a superposition of forward and backward shifts. Since the commonly used trit flip channels rely on cyclic one-parameter permutations, we demonstrate that those formulations are particular cases of the one presented here. Finally, in Section \ref{sec:Entanglement}, we use Negativity as a measure to discuss how entanglement behaves under the action of the newly defined channels on bipartite qutrit systems, demonstrating that the presented channels are not equivalent.

\section{Individual Dit-Flip (IDF): pairwise flipping}\label{sec:IDF}

A way to extend the bit-flip to higher dimensional systems is to regard it as an operation between only two elements of the qudit basis, leaving the remaining elements unchanged. As such, this operation must be reflexive so that the probability $p_{ij}$ of flipping $\ket{i}$ to $\ket{j}$ is the same in the opposite direction, $\ket{j}$ to $\ket{i}$, with $p_{ji}=p_{ij}$. Since the bit-flip is defined using the NOT gate, it is desirable to construct the action of the extended channel using a quantum gate. Therefore, we require that the channel can be expressed, in terms of its Kraus operators, as a unitary operator and the identity. That is, given an input state $\rho$, the action of the channel $\Lambda$ is written as
\begin{equation}\label{eq:Lambda(rho)-1}
    \Lambda\qty(\rho) = p_{ij}\vb{F}_{ij}\rho\vb{F}_{ij}^\dagger + (1-p_{ij})\rho\qc \vb{F}_{ij}\in U(d).
\end{equation}

We start by restricting our analysis to qutrits and consider $\qty{\ket{i},\ket{j},\ket{k}}$ a basis in the qutrit state space written in no particular order. Then, the operator we are searching for is \cite{Di2013-Qudits-Xgate-ITF, Pavlidis2021-Xgate-ITF}

\begin{equation}\label{eq:Fij-qutrit}
    \vb{F}_{ij} = \dyad{i}{j} + \dyad{j}{i} + \dyad{k}{k},
\end{equation}
and the corresponding Kraus operators of this trit-flip channel are

\begin{subequations}\label{eq:ITF-Kraus}
    \begin{equation}
        \vb{K}^{ij}_0 = \sqrt{1 - p_{ij}} \; \mathbbm{1}_3,
    \end{equation}
    \begin{equation}
        \vb{K}^{ij}_1 = \sqrt{p_{ij}} \; \vb{F}_{ij}.\label{eq:Kij1-qutrit}
    \end{equation}
\end{subequations}

Given that the Kraus operators are unitary for all $i,j$, it is trivial to show that this channel is unital \cite{King2002AddUnital, Mendl2009Unital}. That is, $\Lambda(\mathbbm{1})=\mathbbm{1}$. It should also be noticed that, although $\vb{F}_{ij}$ is unitary, we have that $\Tr\vb{F}_{ij}=1$, which differs from the qubit case, where $\Tr\sigma_x=0$.

\subsection{Extension to higher dimensional qudit spaces}

To extend the previous operator and channel to an arbitrary-dimensional qudit space, we only need to include the other elements of the basis that are not involved in the flip. Given the basis $\qty{\ket{k}}$ with $\ket{i}$ and $\ket{j}$ involved in the flipping, the corresponding generalization of \eqref{eq:Fij-qutrit} is

\begin{equation}
     \vb{F}_{ij}^{(d)} = \dyad{i}{j} + \dyad{j}{i} +\sum_{k\neq{}i,j}\dyad{k},
\end{equation}
so the Kraus operators are

\begin{subequations}\label{eq: kraus_idf}
    \begin{equation}\label{eq:K0ij-IDF}
        \vb{K}^{(ij)}_0 = \sqrt{1 - p_{ij}}\; \mathbbm{1}_d,
    \end{equation}
    \begin{equation}\label{eq:K1ij-IDF}
        \vb{K}^{(ij)}_{1} = \sqrt{p_{ij}}\;\vb{F}_{ij}^{(d)}. 
    \end{equation}
\end{subequations}

This construction applies to any qudit system, including 2-dimensional qubits. However, that is the only dimension where the Kraus operator $\vb{K}^{(ij)}_{1}$  is proportional to a $SU(d)$ generator. Also, we can notice that $\Tr\vb{F}_{ij}^{(d)}=d-2$, so that only when $d=2$ we have a unitary operator that has also null trace, that is, an element of $su(d)$.

\subsection{An $su(d)$-based Individual Flip channel}

As previously stated, only in the qubit case does the $\vb{F}_{ij}^{(d)}$ operator correspond to one of the $SU(2)$ generators, namely the Pauli matrix $\sigma_x$. Therefore, we can ask ourselves whether we can define a qudit flip operator for higher dimensional systems that is proportional to an element of the corresponding algebra and, with it, an $su(d)$-based Individual Flip channel.

To address this question, we start by focusing on qutrits. For $d=3$, we can write the corresponding algebra using the Gell-Mann matrices \cite{GellMann-PhysRev.125.1067}

\begin{align}\label{eq:Gell-Mann}
    \lambda_1 &= \mqty(0&1&0\\1&0&0\\0&0&0)\qc\lambda_2 = \mqty(0&-i&0\\i&0&0\\0&0&0)\qc
    \lambda_3 = \mqty(1&0&0\\0&-1&0\\0&0&0), \nonumber\\\lambda_4 &= \mqty(0&0&1\\0&0&0\\1&0&0)\qc\lambda_5 = \mqty(0&0&-i\\0&0&0\\i&0&0)\qc
    \lambda_6 = \mqty(0&0&0\\0&0&1\\0&1&0),\nonumber\\ \lambda_7 &= \mqty(0&0&0\\0&0&-i\\0&i&0)\qc
    \lambda_8 =\frac{1}{\sqrt{3}}\mqty(1&0&0\\0&1&0\\0&0&-2).
\end{align}
As generators of $SU(3)$, Gell-Mann developed them as higher dimensional extensions of the Pauli matrices \cite{georgi1999lie}. That is why it is straightforward to identify the three $su(2)$ subalgebras of $su(3)$ and realize that there are three matrices analog to $\sigma_x$, namely $\lambda_1$, $\lambda_4$, and $\lambda_6$.

Given the abovementioned qutrit basis $\qty{\ket{i},\ket{j},\ket{k}}$, we have
\begin{equation}\label{eq:Gammaij}
    \vb*{\Gamma}_{ij}=\dyad{i}{j} + \dyad{j}{i},
\end{equation}
with
\begin{equation}
    \vb*{\Gamma}_{01}=\lambda_1\qc \vb*{\Gamma}_{02}=\lambda_4\qc \vb*{\Gamma}_{12}=\lambda_6.
\end{equation}
Therefore, the corresponding channel is
\begin{equation}\label{eq:Lambda-tilde(rho)}
    \tilde{\Lambda}(\rho) = p_{ij}\, \vb*{\Gamma}_{ij}\rho\vb*{\Gamma}_{ij} +  \vb{T}_{ij}\rho\vb{T}_{ij},
\end{equation}
where 
\begin{equation}\label{eq:Tij}
    \vb{T}_{ij}=\sqrt{1-p_{ij}\,} \vb*{\Gamma}^2_{ij} +\dyad{k}.
\end{equation}

Unlike the previous implementation, the operator performing the flip is no longer unitary, and we no longer have a Kraus operator proportional to the identity since that would not satisfy the closure condition. Nevertheless, given that the Gell-Mann matrices \eqref{eq:Gell-Mann} were developed as extensions of the Pauli matrices, we regain a null trace.

It is worth noticing that the individual channel is non-unital and that the action of this channel, with $p=1$, over a pure state that includes a basis element not involved in the flip will yield a mixed state, unlike our previous implementation. Nevertheless, if we limit ourselves to the subspace spanned by $\qty{\ket{i},\ket{j}}$,  this quantum channel acts identically to a bit-flip, as expected. 

We call this quantum channel the \textit{su(3)-based Individual Trit Flip} which we can generalize to the family of \textit{su($d$)-based Individual Dit Flip} channels. For higher dimensions, the procedure used by Gell-Mann to construct the matrices in \eqref{eq:Gell-Mann} can be extended to the defining representations of $SU(d)$ so that a realization of $su(d)$ exists in terms of the Generalized Gell-Mann matrices \cite{Bertlmann_2008}. In such case, on a $d$-dimensional system, we can identify $\vb*{\Gamma}_{ij}$ as a generator of $SU(d)$, and from there, define an operator $\vb{T}^{(d)}_{ij}$ for a flip between $\ket{i}$ and $\ket{j}$, analogously to \eqref{eq:Gammaij} and \eqref{eq:Tij}. In such case, the $su(d)$-based flip channel has the following Kraus operators:

\begin{subequations}\label{eq:su(d)IDF-kraus}
    \begin{equation}\label{eq:su(d)IDF-Tij(d)}
        \widetilde{\vb{K}}_0^{(ij)} = \vb{T}^{(d)}_{ij} = \sqrt{1-p_{ij}\,} \vb*{\Gamma}^2_{ij} +\sum_{k\neq{}i,j}\dyad{k},
    \end{equation}
    \begin{equation}\label{eq:su(d)IDF-kraus1}
        \widetilde{\vb{K}}_1^{(ij)} = \sqrt{p_{ij}\,}\;\vb*{\Gamma}_{ij}.
    \end{equation}
\end{subequations}

\section{$su(d)$-based Full Dit Flip channel}\label{sec:su(d)-IDF}

We can define a general flip channel where every interchange $\ket{i} \leftrightarrow \ket{j}$ happens, with probability $p_{ij}$, in a single operation. The operators responsible for applying those flips are like the one defined in Eq. \eqref{eq:su(d)IDF-kraus1} Therefore, this general channel involves all the ${d(d-1)}/{2}$ operators $\widetilde{\vb{K}}_1^{(ij)}$, which in this context we relabel as 

\begin{equation}
    \widetilde{\vb{K}}_{ij}^{(d)} = \sqrt{p_{ij}}\;\vb*{\Gamma}_{ij}\qc 0 \le i < j \le d-1.
\end{equation}

To complete constructing the channel, we might want to consider  including the ${d(d-1)}/{2}$ operators  $\vb{T}^{(d)}_{ij}$ defined in Eq. \eqref{eq:su(d)IDF-Tij(d)}. Nevertheless, the set $\qty{\widetilde{\vb{K}}_{ij}^{(d)}, \vb{T}_{ij}^{(d)}}$ does not satisfy the closure relation and a different approach is required. 

Since, by definition, we must have that for $p_{ij}=0$ for all $0 \le i < j \le d-1$, the transformation performed by the channel is the identity, we propose $\widetilde{\vb{K}}_0$
\begin{equation}
    \widetilde{\vb{K}}_0^{(d)} =\sum_{i=0}^{d-1}\,f\qty(p_{ij})\dyad{i}\qc f\qty(p_{ij}=0)=1.
\end{equation}
to complete the set of Kraus operators. Along with the set $\qty{\widetilde{\vb{K}}_{ij}^{(d)}}$ defined beforehand, they must satisfy the closure relation. Therefore, fter some calculations, we can determine that the operator is

\begin{equation}
    \widetilde{\vb{K}}_{0}^{(d)} = \sum^{d-1}_{i=0} \sqrt{1-\sum_{k\ne i} p_{ik}} \; \dyad{i}{i}.
\end{equation}

With this set of Kraus operators, the result of applying this quantum channel to a qudit state $\rho$ is
\begin{equation}
    {\tilde{\Lambda}}^{(d)}(\rho) = \widetilde{\vb{K}}_0^{(d)}\rho \widetilde{\vb{K}}_0^{(d)} + \sum_{i=0} ^{d-2}\sum_{j=i+1}^{d-1} \widetilde{\vb{K}}_{ij}^{(d)}\,\rho\,\widetilde{\vb{K}}_{ij}^{(d)}.
\end{equation}

Contrary to the individual case, the $su(d)$-based full dit flip channel defined above is no longer non-unital, as one can show that ${\tilde{\Lambda}}^{(d)}(\mathbbm{1})=\mathbbm{1}$.

\subsection{Full Dit-Flip channel}

As we have just done with the $su(d)$-based Individual Dit Flip, we can check what happens when we create a quantum channel that flips all dits, using the IDF channel defined in Eq. \eqref{eq: kraus_idf}. Contrary to what happens when extending the $su(d)$-based IDF to include all possible flip pairs, the process is straightforward by considering the set $\qty{{\vb{K}}_{0}^{(d)},  {\vb{K}}_{ij}^{(d)}}$, where 
\begin{subequations}\label{eq:K0ijKij-CompleteIDF}
    \begin{equation}\label{eq:K0-CompleteIDF}
        {\vb{K}}_0^{(d)} = \sqrt{1-\sum_{i=0}^{d-2}\sum_{j = i + 1}^{d-1} p_{ij}\,} \; \mathbbm{1}_d,
    \end{equation}
    \begin{equation}
         {\vb{K}}_{ij}^{(d)} = \sqrt{p_{ij}\,}\;\vb{F}_{ij}^{(d)}\qc 0 \le i < j \le d-1.
    \end{equation}
\end{subequations}
The resulting state after applying these Kraus operators, after some simple algebraic manipulation, is given by
\begin{equation}
\begin{aligned}
    {\Lambda}^{(d)}(\rho) = &\qty(1-\sum_{i=0}^{d-2}\sum_{j = i + 1}^{d-1} p_{ij})\,\rho\\
    & \qq{}+ \sum_{i=0}^{d-2}\sum_{j = i + 1}^{d-1} {\vb{K}}_{ij}^{(d)}\,\rho\,\qty({\vb{K}}_{ij}^{(d)})^\dagger,
\end{aligned}
\end{equation}   
and remains unital as expected.

\section{Shift dit-flip (SDF): successor and predecessor dit-flip}\label{sec:SDF}

As mentioned before, we can interpret the Dit-Flip as an operation that takes an element of the ordered basis and returns the next or the previous one, while considering the ordered basis as cyclic. However, if we analyze the action of the Dit-Flips we defined in earlier sections, it should be clear that such channels do not act in this way. For $d>2$, shifting and flipping are non-equivalent operations. To highlight this difference, we define a new type of channel, starting with the qutrit case.

Given the standard qutrit computational basis $\qty{\ket{0},\ket{1},\ket{2}}$, the element that follows $\ket{i}$ is $\ket{i+1}$, with $\ket{0}$ following $\ket{2}$, and the element that precedes $\ket{i}$ is $\ket{i-1}$, with $\ket{2}$ preceding $\ket{0}$. In this case, the action involves all elements of the basis \cite{Patera1988PauliGates-Qudits-Shift, Wang2020Qudits-Xgate-F}.

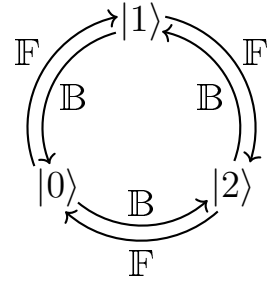
\begin{figure}\begin{center}
\begin{tikzpicture}
    \node at (2.9,0.2) {\Large $\ket{0}$};
    \node at (2.5,2) {\Large $\vb{F}$};
    \node at (3.1,1.4) {\Large $\vb{B}$};
    \draw[->,  thick] ($(4,1) + (200:15mm)$) arc (200:102:15mm);
    \draw[<-,  thick] ($(4,1) + (202:13mm)$) arc (202:105:13mm);
    \node at (4,2.4) {\Large $\ket{1}$};
    \node at (5.5,2) {\Large $\vb{F}$};
    \node at (4.9,1.4) {\Large $\vb{B}$};
    \draw[->,  thick] ($(4,1) + (78:15mm)$) arc (78:-18:15mm);
    \draw[<-,  thick] ($(4,1) + (78:13mm)$) arc (78:-20:13mm);
    \node at (5.2,0.2) {\Large $\ket{2}$};
    \draw[<-,  thick] ($(4,1) + (228:15mm)$) arc (228:312:15mm);
    \draw[->,  thick] ($(4,1) + (230:13mm)$) arc (230:312:13mm);
    \node at (4,-.8) {\Large $\vb{F}$};
    \node at (4,0) {\Large $\vb{B}$};
    \end{tikzpicture}
    \caption{Action of the $\vb{F}$ and $\vb{B}$ operators on the qutrit computational basis.}
    \label{fig:ShiftTritFlip-Diagram}
\end{center}\end{figure}

Both these operations can be synthesized into a two-parameter quantum channel. To construct such an operation, let us first define the forward $\vb{F}$ and backward $\vb{B}$ operators as:

\begin{subequations}\label{eq:F&B}
\begin{equation}
    \vb{F}= \mqty(
        0 & 0 & 1 \\
        1 & 0 & 0 \\
        0 & 1 & 0),
\end{equation}
\begin{equation}
    \vb{B}= \mqty(
        0 & 1 & 0 \\
        0 & 0 & 1 \\
        1 & 0 & 0),
\end{equation}
\end{subequations}
where $\vb{F},\vb{B}\in{}SU(3)$. These unitary operators satisfy the following relations

\begin{subequations}\label{eq:Prop-BF}
    \begin{equation}
    \vb{F}^\dagger=\vb{B}
    \end{equation}
    \begin{equation}
    \vb{F}^2=\vb{B}    
    \end{equation}
    \begin{equation}
        \vb{B}^2=\vb{F}
    \end{equation}
    \begin{equation}
        \qty[\vb{B},\vb{F}]=0
    \end{equation}
\end{subequations}
In Figure \ref{fig:ShiftTritFlip-Diagram}, we present a schematic visualization of the action of these operators. 

Let $p$ be the probability of the flip happening at all, $f$ the \textit{forward coefficient}, and $b$ the \textit{backward coefficient}, such that $b, f \in [0,1]$. The quantum channel's action is

\begin{equation}\label{eq:Lambda_s(rho)}
    \Lambda_s(\rho) = (1-p) \rho + p\left(f\vb{F} \rho \vb{B} + b\vb{B}\rho \vb{F}\right),
\end{equation}
and the associated Kraus operators are the following:

\begin{subequations}\label{eq:Kraus_STF}
\begin{equation}
    \vb{K}_0 = \sqrt{1-p\,}\;\mathbbm{1}_3,\label{eq:Kraus_STF-K0}
\end{equation}
\begin{equation}
    \vb{K}_1 = \sqrt{pf\,} \; \vb{F},\label{eq:Kraus_STF-K1}
\end{equation}
\begin{equation}
    \vb{K}_2 = \sqrt{p\,b\,} \; \vb{B}.\label{eq:Kraus_STF-K2}
\end{equation}
\end{subequations}
The closure relation imposes that $b + f = 1$. If we have $f=1$, we have a successor operation and the equivalent of the X gate for the qutrit system introduced by Lawrence when discussing mutually unbiased bases (MUBs) \cite{PhysRevA.70.012302} and later extended to arbitrary d-dimensional gates by Gao et al. \cite{PhysRevA.99.023825}. Analogously, with $b=1$, the quantum channel will be equivalent to a predecessor operation. 

For this channel to become a quantum gate when $p=1$, we require that either $f$ or $b$ equal to unity. As discussed when we introduced the $\vb{F}$ and $\vb{B}$ operators, these operators are in $SU(3)$. Moreover, as is the case for the Pauli matrices, these operators are traceless as well as unitary. 

Nevertheless, we have to mention that the $\vb{F}$ and $\vb{B}$ operators do not belong to the $su(3)$ algebra since these are not skew-Hermitian matrices, as required for elements in $su(d)$ (See Proposition 3.24 in  \cite{Hall2015LieGroupsAlgRep}), or symmetric, as is the case for the qubit X gate, with $i\sigma_x$ belonging to $su(2)$. This difference highlights an intrinsic limitation imposed on the extension of Pauli-like gates to higher dimensions: the need to choose which characteristics one intends to preserve in these quantum gates, as there are no unitary, traceless, and (skew-)symmetric $d\times{d}$ matrices for $d\geq3$.

\subsection{Extension to general qudit spaces}

To extend this channel to higher-dimensional systems, we define the successor $\mathcal{F}: \mathbbm{N} \to \mathbbm{N}$ and predecessor $\mathcal{B}: \mathbbm{N} \to \mathbbm{N}$ functions as

\begin{subequations}\label{eq:Bi&Fi-functions}
    \begin{equation}
        \mathcal{B}(i) = \begin{cases}
            i - 1 & i > 0 \\
            d - 1 & i = 0
        \end{cases},
    \end{equation}
    \begin{equation}
        \mathcal{F}(i) = \begin{cases}
            i + 1 & i < d - 1 \\
            0 & i = d - 1
        \end{cases},
    \end{equation}
\end{subequations}
where the computational qudit basis is $\qty{\ket{i}:i=0,\ldots,d-1}$.

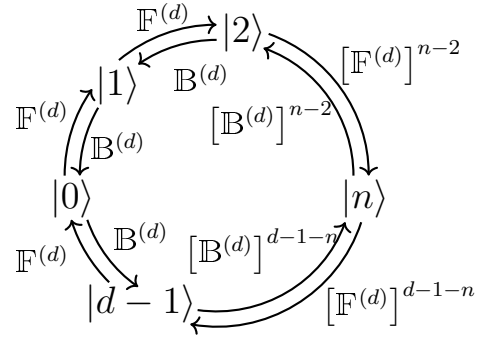
\begin{figure}[ht!]\begin{center}
\begin{tikzpicture}
    \node at (2.1,0.2) {\Large $\ket{0}$};
    \node at (1.9,1.65) {\large $\vb{F}_{(d)}$};
    \node at (2.9,1.2) {\large $\vb{B}_{(d)}$};
    \draw[->,  thick] ($(4,.5) + (180:20mm)$) arc (180:115:20mm);
    \draw[<-,  thick] ($(4,.5) + (180:18mm)$) arc (180:120:18mm);
    \node at (3.5,2.2) {\Large $\ket{1}$};
    \draw[->, densely dotted, thick] ($(4,.5) + (95:20mm)$) arc (95:30:20mm);
    \draw[<-, densely dotted, thick] ($(4,.5) + (95:18mm)$) arc (95:30:18mm);
    \node at (5.9,1) {\Large $\ket{n}$};
    \draw[<-, densely dotted, thick] ($(4,.5) + (5:20mm)$) arc (5:-60:20mm);
    \draw[->, densely dotted, thick] ($(4,.5) + (5:18mm)$) arc (5:-60:18mm);
    \node at (4.1,-1.2) {\Large $\ket{d-1}$};
    \draw[<-, thick] ($(4,.5) + (197:20mm)$) arc (197:250:20mm);
    \draw[->, thick] ($(4,.5) + (199:18mm)$) arc (199:250:18mm);
    \node at (2.1,-1.) {\large $\vb{F}_{(d)}$};
    \node at (3.,-0.5) {\large $\vb{B}_{(d)}$};
    \end{tikzpicture}
    \caption{Action of the $\vb{F}_{(d)}$ and $\vb{B}_{(d)}$ operators on the qudit computational basis.}
    \label{fig:ShiftDitFlip-Diagram}
\end{center}\end{figure}

With these functions, we can now generalize \eqref{eq:Lambda_s(rho)} to higher dimensions by defining the corresponding forward and backward operators as $\vb{F}_{(d)}$ and $\vb{B}_{(d)}$, respectively.
\begin{subequations}\label{eq:F(d)&B(d)}
        \begin{equation}
             \vb{F}_{(d)}= \sum^{d-1}_{i=0} \dyad{\mathcal{F}(i)}{i},
        \end{equation}
        \begin{equation}
            \vb{B}_{(d)} = \sum^{d-1}_{i=0} \dyad{\mathcal{B}(i)}{i}.
        \end{equation}
\end{subequations}
The action of these operators on the elements of the qudit computational basis is presented schematically in Figure \ref{fig:ShiftDitFlip-Diagram}. 

Since \eqref{eq:F&B} can be written using the $\mathcal{B}$ and $\mathcal{F}$ functions \eqref{eq:Bi&Fi-functions} as $\vb{F}=\vb{F}_{(3)}$ and $\vb{B}=\vb{B}_{(3)}$ respectively, it is straightforward to realize that the previous operators belong to $SU(d)$ and fulfill the following relations: 
\begin{subequations}
    \begin{equation}
        \vb{F}_{(d)}^\dagger = \vb{B}_{(d)},
    \end{equation}
    \begin{equation}
        \vb{F}_{(d)}^{d-1} = \vb{B}_{(d)},
    \end{equation}
    \begin{equation}
        \vb{B}_{(d)}^{d-1} = \vb{F}_{(d)},
    \end{equation}
    \begin{equation}
        \qty[\vb{B}_{(d)}, \vb{F}_{(d)}] = 0.
    \end{equation}
\end{subequations}

From the above properties and since $\vb{F}_{(d)}\ket{0} =\ket{1}$, we can readily notice that we can relate any two basis vectors $\ket{i}$ and $\ket{j}$ using the above-defined operators
\begin{equation}
    \ket{i}= \vb{F}_{(d)}^{j-i}\ket{j} = \vb{B}_{(d)}^{d-i+j}\ket{j}.
\end{equation}
Therefore, we have that
\begin{equation}
    \vb{F}_{(d)}^n= \vb{B}_{(d)}^{d-n}.
\end{equation}

The Kraus operators for the Shift Dit Flip channel, the higher-dimensional generalization of \eqref{eq:Lambda_s(rho)}, are
\begin{subequations}\label{eq:Kraus_SDF}
        \begin{equation}\label{eq:Kraus_SDF-K0}
            \vb{K}_0^{(d)} = \sqrt{1-p\,}\;\mathbbm{1}_d,
        \end{equation}
        \begin{equation}
            \vb{K}_1^{(d)} =  \sqrt{pf\,}\; \vb{F}_{(d)},
        \end{equation}
        \begin{equation}
            \vb{K}_2^{(d)} = \sqrt{p\,b\,}\; \vb{B}_{(d)}.
        \end{equation}
\end{subequations}
where the closure relations requires that $f+b=1$.

\subsection{Shuffled Shift Dit Channel}

The qudit computational basis can have $(d-1)!$ different orderings. For $d\geq4$, not all such orderings are attainable by applying the $\vb{F}_{(d)}$ or $\vb{B}_{(d)}$ operators a given number of times to the original basis. Therefore, we can consider functions beyond the predecessor and successor ones defined beforehand by first reordering the qudit computational basis.

In particular, suppose that $D = \{0, 1, \dots, d-1\}$. Then, there exist $(d-1)!$ unique bijective functions $F_i: D \to D$ for every possible ordering in the system, such that if $B=\qty{\ket{0}, \ket{1},\ldots, \ket{d-1}}$ is the qudit standard computational basis, $F_i(B)$ is a reordering of it, such that $F_i(B)$ and $F_j(B)$ are bases with different orderings if $j\ne{}i$. Let $f_i$ be $(d-1)!$ real non-negative parameters. We can define the following Kraus operators for a channel representing a shift for every order.

\begin{subequations}
        \begin{equation}
            \vb{K}_0 = \sqrt{1-p}\;\mathbbm{1}_d,
        \end{equation}
        \begin{equation}
            \vb{K}_j = \sqrt{pf_j} \sum^{d-1}_{i = 0} \dyad{F_j(i)}{i} \qc j = 1, 2, \dots, (d-1)!
        \end{equation}
\end{subequations}

After some straightforward calculations and algebraic manipulation, we can verify that the Kraus operators defined obey the closure relation when
\begin{equation}
    \sum_{j=0}^{(d-1)!} f_i = 1.
\end{equation}

\subsection{Shift Trit Channel and the widely used one-parameter Trit-Flip channels}\label{sec:GSDF}

As far as we know, the trit-flip channels most commonly used in the literature are one-parameter cyclic permutations that are directly related to the previously discussed Shift Trit Channel. This formulation naturally arises when considering the X gate as the operation that projects a state into an orthogonal one,
\begin{equation}
    \mathbbm{X}:\; \ket{i}\to \cos\theta\ket{i+1}+\sin\theta\ket{i-1}.
\end{equation}
Such an operation is not unitary, but from our previous discussion, it is straightforward to write it in terms of unitary operators as $\cos\theta\,\vb{F}+\sin\theta\,\vb{B}$. For $\theta=\pi/4$, the operator is proportional to $\vb{F} + \vb{B}$, leading to the one-parameter cyclic version of this channel.

As mentioned earlier, if we have $f=1$ in \eqref{eq:Lambda_s(rho)}, then the shifting part of the channel is a purely successor operation. The $\vb{F}$ operator is then the qutrit equivalent of the X gate introduced by Lawrence when discussing MUBs \cite{PhysRevA.70.012302}. If instead, we have $f = b = 1/2$, we would obtain a channel proposed as the trit-flip on works studying qutrit entanglement \cite{Wei_2012, jiang2024joint}.

Another one-parameter cyclic channel directly related to our proposed shift-based channel can be found in Doustimotlagh et al. \cite{doustimotlagh2015quantum} and Wei et al.\cite{wei2013geometric}, to name a few. In this case, we rescale the $p$ parameter as $p\longrightarrow{}p'=2p/3$, and the forward and backward coefficients become $f=b=1/2$.

This last example leads to considering a slight modification of the proposed Shift Trit Channel \eqref{eq:Kraus_STF} that is not a mere rescaling and fixing of the parameters. Let $t$ be a scaling factor so that $0 \le t \le 1$. By introducing a new $\vb{K}'_0$ operator
\begin{equation}
    \vb{K}'_0 = \sqrt{1 - tp}\; \mathbbm{1}_3
\end{equation}
and maintaining the definition for $\vb{K}_1$ \eqref{eq:Kraus_STF-K1} and $\vb{K}_2$ \eqref{eq:Kraus_STF-K2}, we have a three-parameter channel instead of the original two-parameter Shift Trit Channel \eqref{eq:Kraus_STF}. With these operators, the closure relation requires that
\begin{equation}\label{eq:DSTF-bft}
    f+b=t.
\end{equation}

For $p=1$, the fraction of the original state that remains unchanged is $1-t$, while $t$ is the proportion that gets shifted. For $t = 0$, this channel reduces to the identity for any $p$ since there is no transformation, since \eqref{eq:DSTF-bft} requires that $b=f=0$. For $t=1$, we recover the proposed shift-based flip studied in the previous section, which is therefore a special case of this formulation. 

It is straightforward to generalize this channel to higher dimensions by using the $\vb{K}^{(d)}_1$ and $\vb{K}^{(d)}_2$ operators \eqref{eq:Kraus_SDF}, and $\mathbbm{1}$ in $\vb{K}'_0$ being of the appropriate dimension. Given the above discussion and the character acquired by the $t$ parameter for $p=1$, we label this formulation as the \textit{Damped Shift Dit Flip} (DSDF) channel.

\section{Entanglement under the Trit-Flip channels}\label{sec:Entanglement}

We are now interested in how the quantum channels presented so far modify entanglement. To do this, we focus on qubit-qutrit and 2-qutrit extensions of Werner states \cite{Werner-PhysRevA.40.4277}. As is common in the literature for such higher-dimensional Werner states, we consider a statistical mix of a maximally entangled state and a maximally mixed state,
\begin{equation}\label{eq:Werner-qudit}
    \rho_w^{AB}= a\dyad{\Psi_{ME}} + \frac{1-a}{d_A\,d_B}\,\mathbbm{1}_{d_A\,d_B},
\end{equation}
where $d_A$ and $d_B$ are the dimensions of the $A$ and $B$ subsystems, respectively.

As for the entanglement measure analyzed in this section, we focus on Negativity \cite{Vidal_2002, Negativity-qudits}. For a given state $\rho^{AB}$ written as
\begin{equation}\label{eq:rho_AB-ijmn}
    \rho_{AB}=\sum_{i,j,n,m}c_{ij}^{mn}\dyad{ij}{mn},
\end{equation}
its partial transpose is defined as
\begin{equation}\label{eq:rho_AB-pt}
    \rho_{AB}^{\,pt}=\sum_{i,j,n,m}c_{ij}^{mn}\dyad{in}{mj},
\end{equation}
whose eigenvalues we label as $\lambda_i$. The Negativity, derived from the PPT criterion \cite{Horodecki-PPT}, is defined as \cite{Negativity-qudits}
\begin{equation}
    \mathcal{N}\qty(\rho_{AB})\equiv \frac{1}{d-1}\qty\Big[\norm{\rho_{AB}^{\,pt}}_1-1],
\end{equation}
where 
\begin{equation}
    \norm{\rho_{AB}^{\,pt}}_1 = \Tr\sqrt{\qty(\rho_{AB}^{\,pt})^\dagger\rho_{AB}^{\,pt}\;}
\end{equation}
is the trace norm of $\rho_{AB}^{\,pt}$ and $d=\min\qty(d_A,d_B)$. Alternatively, we can use the equivalent definition involving all eigenvalues of $\rho_{AB}^{\,pt}$:
\begin{equation}\label{eq:N-lambda_i}
    \mathcal{N}\qty(\rho_{AB})=\frac{1}{2}\,\sum_i\qty(\qty|\lambda_i|-\lambda_i).
\end{equation}

As an entanglement measure, $\mathcal{N}\qty(\rho_{AB})\neq0$ is a necessary and sufficient condition for entanglement in 2-qubit and qubit-qutrit systems. For higher-dimensional systems, this condition is only sufficient since there can be entangled states with null negativity. Therefore, even though this measure does not allow us to study the separability of 2-qutrit states, it still allows us to analyze how entanglement behaves. In what follows, we apply the channels over one qutrit subsystem, leaving the other subsystem unaltered.

\subsection{Qubit-Qutrit}

\begin{figure}[t]
    \centering
    \includegraphics[width=\linewidth]{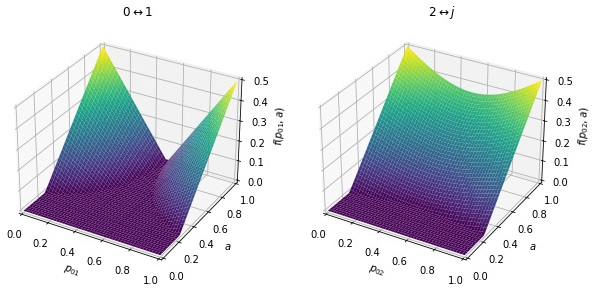}
    \caption{Negativity as a function of the state parameter $a$ and the channel parameter $p_{ij}$ by the $\ket{i}\leftrightarrow\ket{j}$ ITF channel on a qubit-qutrit Werner state.}
    \label{fig:itf_23}
\end{figure}

To define the Werner state that we study as mixed qubit-qutrit states, we use
\begin{equation}\label{eq:Phi+23}
    \ket{\Phi^+_{23}} = \frac{1}{\sqrt{2}}\qty( \ket{00} + \ket{12})
\end{equation}
as $\ket{\Psi_{ME}}$, the maximally entangled one \cite{MaxEntStates-QubitQutrit, Hedemann_2022} in Eq. \eqref{eq:Werner-qudit} and start by applying the ITF channel \eqref{eq:ITF-Kraus} for $\ket{0}\leftrightarrow\ket{1}$. In Figure \ref{fig:itf_23}, we present the surfaces generated by the Negativity as a function of the state parameter $a$ and the channel parameter $p_{ij}$. Although we only plot the results for a particular flip, similar surfaces result when studying the remaining flips.

As can be seen, we have two distinct behaviors depending on the presence of the element not involved in the flip. Both Negativities are symmetric, and it is noticeable how the entanglement is reduced drastically near $p_{01} = 1/2$. Such behavior for the $\ket{0} \leftrightarrow \ket{1}$ flip is a direct consequence of the usage of $\ket{\Psi_{ME}} = \ket{\Phi^+_{23}}$ and how this state becomes a separable state around $p_{01} = 1/2$.

\begin{figure}[ht!]
    \centering
    \includegraphics[width=\linewidth]{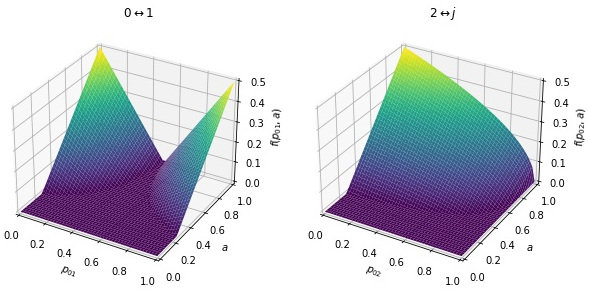}
    \caption{Negativity as a function of the state parameter $a$ and the channel parameter $p_{ij}$ by the $\ket{i}\leftrightarrow\ket{j}$ $su(d)$-based ITF channel on a qubit-qutrit Werner state.}
    \label{fig:itf2_23}
\end{figure}

As for the $su(d)$-based ITF channel \eqref{eq:Lambda-tilde(rho)}, we present the surfaces generated by the Negativity \cite{Vidal_2002} as a function of the state parameter $a$ and the channel parameter $p_{ij}$ in Figure \ref{fig:itf2_23}. For this formulation, the behavior is different and non-symmetrical for the flips involving $\ket{2}$, the element not present in $\ket{\Phi^+_{23}}$. This difference happens because, for $p_{2i} = 1$, the resulting state becomes separable \cite{PhysRevA.78.024303}.

\begin{figure}[ht!]
    \centering
    \includegraphics[width=\linewidth]{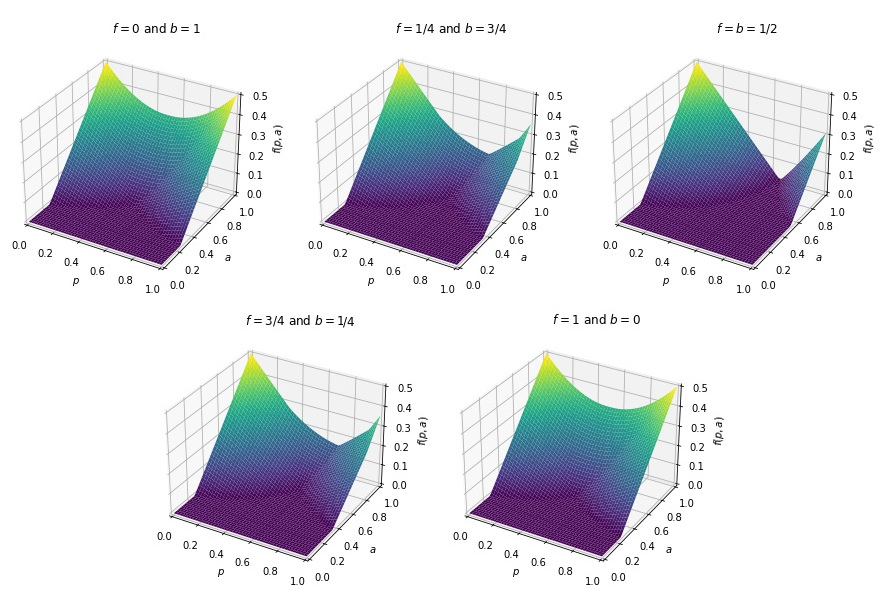}
    \caption{Negativity as a function of the qubit-qutrit Werner state parameter $a$ and the STF channel parameter $p$ for various values of $f$ and $b$}
    \label{fig:stf_23}
\end{figure}

Next, we consider the Shift Trit Flip channel discussed in Section \ref{sec:SDF}. We present the different behaviors of the Negativity of the qubit-qutrit Werner state \eqref{eq:Werner-qudit} under the action of this channel in Figure \ref{fig:stf_23}. Here, it is noticeable how the forward and backward operations influence the state's Negativity. Also, we can see how interchanging $b$ and $f$  does not affect it. Moreover, we note that the further they get close to $1/2$, the sharper the loss in entanglement is.

\begin{figure}[ht!]
    \centering
    \includegraphics[width=\linewidth]{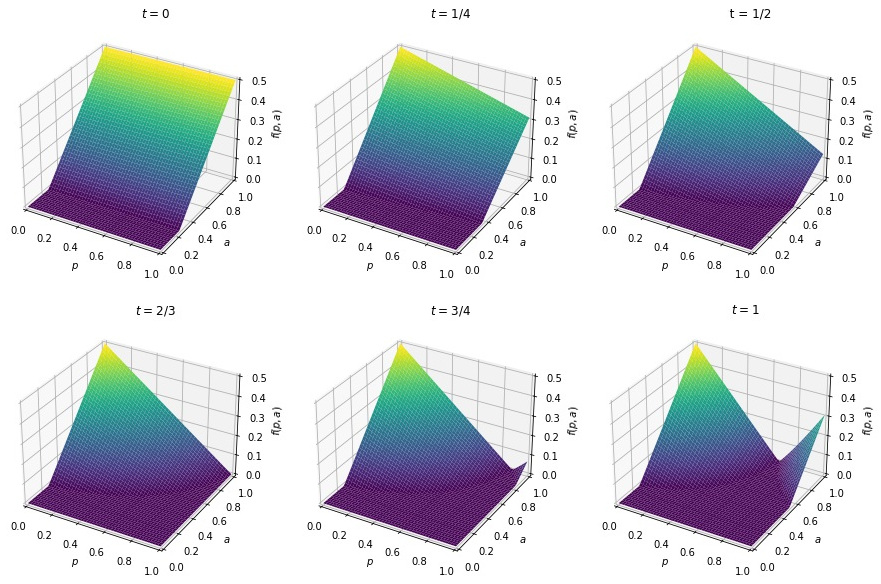}
    \caption{Negativity of a qubit-qutrit Werner state under the action of the Damped Shift Trit Flip channel for various values of $t$ and $b = f = t/2$.}
    \label{fig:gstf_23}
\end{figure}

Finally, we consider the Damped Shift Trit Flip Channel discussed in Section \ref{sec:GSDF}. In Figure \ref{fig:gstf_23}, the graphs show how the parameter $t$ influences entanglement for $b = f = t/2$. As the scaling factor $t$ increases, more entanglement is lost. Such behavior is due to the channel, with a higher transformation rate, having a major impact on the maximally entangled state.

\subsection{2-Qutrit}

For the 2-qutrit Werner state \cite{Qutrit-Ent-99}, we use
\begin{equation}
    \ket{\Phi^+_{33}} = \frac{1}{\sqrt{3}}\qty(\ket{00} + \ket{11} + \ket{22})
\end{equation}
as the maximally entangled state $\ket{\Psi_{ME}}$ in \eqref{eq:Werner-qudit}. As with the 2-qubit Werner states, a state like this is highly symmetrical. In particular, it is invariant under subsystem exchange, and we can apply the channel to either subsystem.

\begin{figure}[ht!]
    \centering
    \includegraphics[width=.75\linewidth]{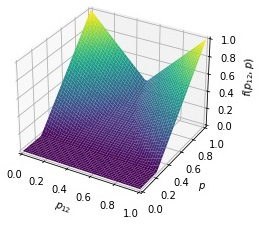}
    \caption{Negativity of the 2-qutrit Werner state as a function of the state parameter $a$ and the  ITF channel parameter $p_{ij}$.}
    \label{fig:itf_33}
\end{figure}

As we did for the qubit-qutrit case, we start by analyzing the ITF channel. In Figure \ref{fig:itf_33}, the graphs show how the Negativity evolves after applying an ITF channel with any pair of flips. Unlike what we obtained for the qubit-qutrit case, there is no entanglement death for any value of the channel's parameter. Since our maximally entangled stated includes all elements of the qutrit basis, the flips do not lead to a separable state.

As was the case for the ITF channel, for the $su(d)$-based one, no matter what flip $\ket{i}\leftrightarrow\ket{j}$ happens, the behavior of the Negativity is the same. The symmetry centered in $p_{ij}=1/2$ observed in the previous formulation is absent in this channel. While the Negativity decreases linearly for $0\le p\le 1/2$, given a fixed value of the state parameter $a$, it stays approximately constant near $1/2$, finally decreasing as it approaches $p=1$.

\begin{figure}[ht!]
    \centering
    \includegraphics[width=.75\linewidth]{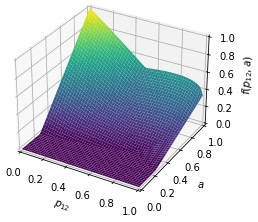}
    \caption{Negativity of the 2-qutrit Werner state as a function of the state parameter $a$ and the  $su(d)$-based ITF channel parameter $p_{ij}$.}
    \label{fig:itf2_33}
\end{figure}

\begin{figure}
    \centering
    \includegraphics[width=\linewidth]{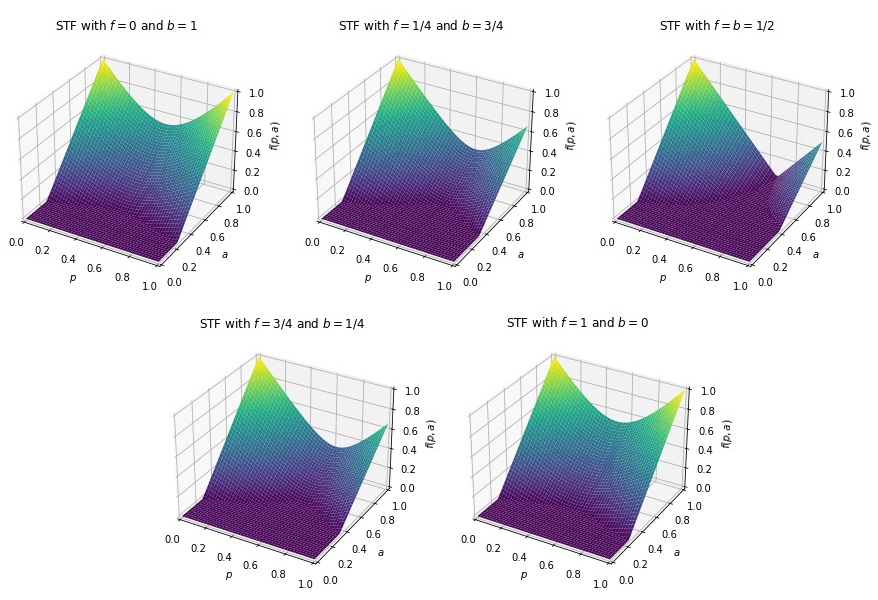}
    \caption{Negativity of the Werner State in a 2-qutrit system under the action of the STF channels for various values of $f$ and $b$.}
    \label{fig:stf_33}
\end{figure}

\begin{figure}
    \centering
    \includegraphics[width=\linewidth]{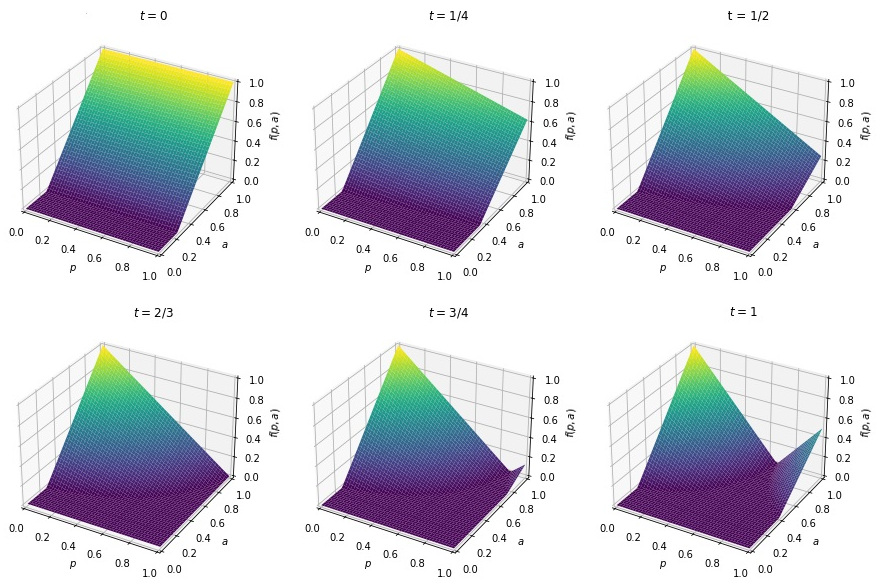}
    \caption{Negativity of a 2-qutrit Werner State under the action of the Damped Shift Trit channel for various values of $t$ and $b = f = t/2$.}
    \label{fig:gstf_33}
\end{figure}

Finally, we present the results obtained for the shift-based flip channels discussed in Section \ref{sec:SDF} in Figures \ref{fig:stf_33} and \ref{fig:gstf_33}. For the STF channel \eqref{eq:Kraus_STF}, the results are similar to the ones obtained for the qubit-qutrit state. Similarly, we can see in Figure \ref{fig:gstf_33} how the value of transformation parameter $t$ influences the total entanglement. The surfaces are exactly the same as before, but the difference lies in the scaling, which stems from the maximum value for the Negativity in 2-qutrit and qubit-qutrit systems.

\section{Summary and conclusions}\label{sec:Summary&Conclusions}

We have extended the formulation of the qubit flip channel to higher- dimensional systems. This process highlights the implications of the equivalence 2+2=2*2, which hides a rich structure that can only be uncovered by analyzing d-dimensional systems. Starting from the next simple system, the qutrit, we formulated various implementations of the qudit flip channel, each representing different actions. 

After presenting the different trit flip channels, we formulated them for higher-dimensional systems, naming them dit flip channels. Although the qutrit formulation already revealed various implementations, when studying higher-dimensional qudits, we found that an additional formulation arose: that of a rearrangement of the qudit computational basis before performing a shift, leading to the Shuffled Shift Dit Flip channel.

We also briefly discussed the mathematical constraints of extending the Pauli X gate and the selection of the characteristics one intends to preserve in such an extension. In particular, when discussing the commonly used cyclic permutation channel, we mention that the \textbf{F} and \textbf{B} operators are not symmetric, unlike the qubit X gate. 

Finally, we applied the various formulations presented in this study to qubit-qutrit and 2-qutrit Werner states and analyzed their Negativity as an entanglement measure. In doing so, we highlighted the distinctiveness of those different versions of the trit-flip channels.

\section*{Acknowledgments}
The authors wish to thank Prof. Dr. Douglas Mundarain, Universidad Católica del Norte, Chile, for his comments and suggestions for the preliminary version of this article. Albrecht would also like to acknowledge the support of the research group GID-30, \emph{Teoría de Campos y Óptica Cuántica}, at the Universidad Simón Bolívar, Venezuela.


\section*{Declarations}

No funding was received to assist with the preparation of this manuscript.

During the preparation of this work, the authors used free web-based AI-assisted proofreader (Grammarly, non-subscription) for improving grammar, spelling, and wording, reviewing and editing the output as needed. The authors take full responsibility for the final content and attest that it represents their own intellectual effort. 

\section*{Data availability statement}
The authors declare that no datasets were generated or analyzed during this study. Therefore, no underlying data are available for this article.

\bibliographystyle{unsrt}
\bibliography{FlippinQudits}

\end{document}